\documentclass[doublecol]{epl2}
\usepackage{amssymb}
\usepackage{graphicx}
\usepackage{dcolumn}
\usepackage{amsmath}
\usepackage{bm}
\usepackage{epsfig}

\newcommand{\be}{\begin{equation}}
\newcommand{\ee}{\end{equation}}

\title{Continuous- and discrete-time Glauber dynamics.
First- and second-order phase transitions
in mean-field Potts models}
\shorttitle{First- and second-order phase transitions
in mean-field Potts models}
\author{M. Ostilli \inst{1,2}, F. Mukhamedov \inst{3}}
\institute{ \inst{1}
Cooperative Association for Internet Data Analysis, 
San Diego Supercomputer Center, UCSD, San Diego, CA\\
\inst{2}
Statistical Mechanics and Complexity Center (SMC), INFM-CNR SMC, Rome, Italy\\
\inst{3} Dept. of Computational \& Theoretical Sciences, Faculty of Science, IIUM,
Kuantan, Malaysia 
} 


\abstract{
As is known, at the Gibbs-Boltzmann equilibrium, the mean-field $q$-state Potts model 
with a ferromagnetic coupling has only a first order phase transition when $q\geq 3$,
while there is no phase transition for an antiferromagnetic coupling.
The same equilibrium is asymptotically reached when one considers the continuous time evolution according
to a Glauber dynamics.
In this paper we show that, when we consider instead the Potts model evolving
according to a discrete-time dynamics, the Gibbs-Boltzmann equilibrium is reached
only when the coupling is ferromagnetic while,  
when the coupling is anti-ferromagnetic,
a period-2 orbit equilibrium is reached and a stable second-order
phase transition in the Ising mean-field universality class sets in for each component of the orbit.
We discuss the implications of this scenario in real-world problems.
}

\begin{document}

\maketitle

\email{ostilli@roma1.infn.it}

\section{Introduction} 
The $q$-state Potts model is one of the most important models in statistical physics \cite{Wu}.
For example, in the limit $q\to 1$, the model
is equivalent to bond-percolation phenomena \cite{Fortuin}.
More in general, the model finds countless applications in computer science,
like coloring a random graph \cite{Zecchina,MezardC},
and extracting communities from real-world networks \cite{Reichardt} (for a review see also \cite{Review}).
It is perhaps the first candidate model when one wants to consider
first-order phase transition phenomena which, as opposed to second-order
transitions, are characterized by a finite jump of the order parameter.
In the mean-field version of the model, for example, where any spin interact with any other spin, 
it is easy to show that,
at the Gibbs-Boltzmann equilibrium,
when the number of states $q>2$, and the coupling is ferromagnetic, a first-order transition takes place
while, for $q=2$, the model is equivalent to the Ising model and only 
a second-order phase transition takes place. 
When the coupling is instead anti-ferromagnetic, no phase transition sets in, for any $q$,
and the system remains in its trivial symmetric solution.
The same scenario is found when one uses a dynamical approach like the 
continuous-time Glauber dynamics \cite{Glauber}: after an initial transient time
the systems reaches the Gibbs-Boltzmann equilibrium, 
in both the ferromagnetic and anti-ferromagnetic case. 

{
In this paper, after reviewing the mean-field Potts model at equilibrium,
and along the continuous-time Glauber dynamics,
we introduce a discrete-time 
{``Glauber''} dynamics. 
We then show that, when the coupling
is ferromagnetic, the differences between the continuous- and the discrete-time dynamics are irrelevant.
However, when the coupling is anti-ferromagnetic, 
along the discrete-time dynamics the system
reaches a dynamical stability by oscillating regularly between two values.
}
In other words, when the dynamics is discrete and the coupling anti-ferromagnetic, 
the system never reaches a point-like equilibrium, 
but rather a period-2 stable trajectory. 
{Moreover}, there exists a critical temperature 
{
where
}
the system undergoes a second-order phase
transition in the mean-field Ising universality class, for any $q\geq 2$. 

Perhaps, the reader familiar with the Ising model ($q=2$) will not find surprising
that in the anti-ferromagnetic case a discrete-time dynamics
can produce stable finite oscillations because, at any two dynamic iterations, 
the effective coupling becomes ferromagnetic. 
{Ising models with a discrete-time dynamics were studied long ago in the context of 2-state neural networks and cellular automata and, later,
also in the random field Ising model \cite{Synchro}}. 
However, in the Potts case ($q>2$),
we find out also the non intuitive feature that the nature of the phase transition
changes dramatically: the fact that the effective coupling is ferromagnetic
does not return a first-order phase transition, but a second-order
phase transition. 

{
Continuous phase transitions in the 
Potts model with $q\geq 3$ were observed in complex networks with power
law exponent $\gamma\leq 3$ \cite{Review}, and also on the complete graph
in the microcanonical ensemble (as opposed to the canonical, or Gibbs-Boltzmann, ensemble) 
\cite{Costeniuc}. However, the dynamical nature of the continuous transition we 
present in this paper is totally different from the above equilibrium cases.
}
We stress also that the discrete-time dynamics we shall focus on, is not meant 
as an approximation of the continuous-time dynamics (as is instead
usually done for practical simulations). There are in fact infinitely many
remarkable examples where the dynamics is intrinsically discrete.
Whereas only a continuous-time dynamics can represent some description
of a system of physical particles each other interacting via a physical medium, 
a discrete-time dynamics can represent a system of agents 
which interact via, \textit{e.g.}, exchange of information
taking place at discrete random times, as in fact occurs in the actual world,
especially in social or economical contexts, but also in ecosystems.
As we will show, however, unlike the continuous case, 
in order to be well defined at all times,
the discrete-time master equation
requires imposing a bound on how fast free spins can change status
(an external parameter independent from the Hamiltonian parameters).

The paper is organized as follows.
In the next Section we
review the mean-field Potts model and discuss
the existence of an unstable second-order phase transition for which
we will find later a stable counterpart.
In Sec. III we introduce the continuous- and discrete-time Glauber dynamics
and discuss in detail their common points and differences.
We then calculate the critical point and the critical
behavior of the discrete time-dynamics and compare with numerical analysis.
Conclusions are then draw.

\section{The mean-field Potts model}
\begin{figure}[thb]
\includegraphics[scale=0.35]{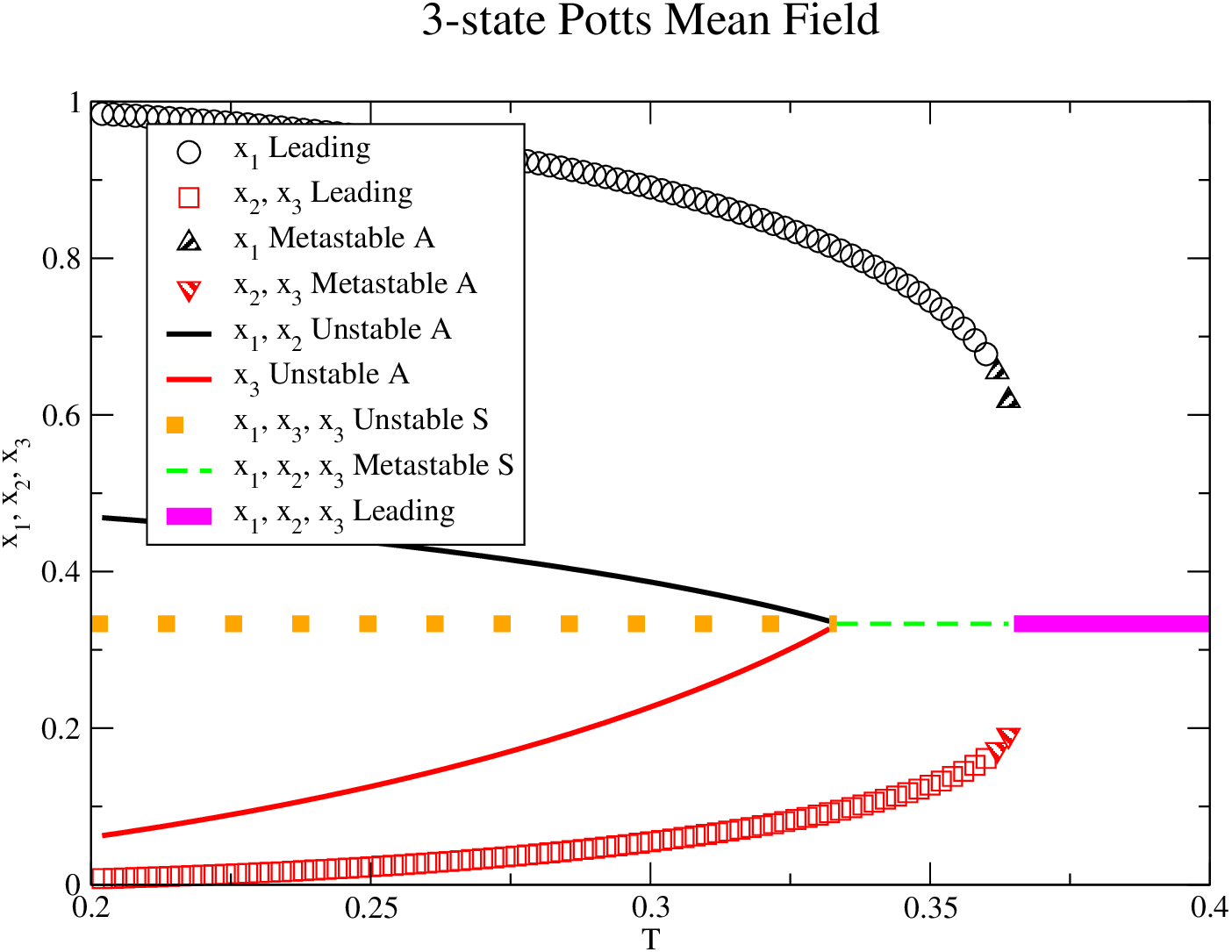}
\caption{(Color online) Magnetizations at equilibrium for the case $q=3$ and $J=1$.
The first-order phase transition takes place at the critical temperature given
by Eq. (\ref{tc}) which gives $T^{(\mathrm{F.O.})}_c=0.36067$, while the critical temperature of 
the unstable second-order phase transition is given by Eq. (\ref{tc1}), which gives $T^{(\mathrm{S.O.~unstable})}_c=1/3$.  
{
In the figure: ``Leading'' stands for the thermodynamic stable state, \textit{i.e.}, having the lower
free-energy, ``Metastable'' stands for the stable state having the higher   
free-energy, while ``Unstable S'' and ``Unstable A'' stand for the 
unstable states having the symmetry $x_1=x_2=x_3=1/3$ or not, respectively.
}
The Leading states lie on the subspaces $x_{i_1}>x_{i_2}=x_{i_3}$
(in this figure only the case $x_1>x_2=x_3$ is shown),
while the Unstable A states lie on the subspaces $x_{i_1}=x_{i_2}>x_{i_3}$
(in this figure only the case $x_1=x_2>x_3$ is shown), where
$i_1,i_2,i_3$, is any permutation of the set of indices $\left\{1,2,3\right\}$ 
and $x_{i_1}+x_{i_2}+x_{i_3}=1$.
Note that the asymptotic values of the magnetizations toward $T=0$ 
are $x_{i_1}=1,~x_{i_2}=x_{i_3}=0$ and $x_{i_1}=x_{i_2}=1/2,~x_{i_3}=0$, for the Leading
and Unstable A states, respectively.  
}
\label{fig1}
\end{figure}

We now briefly review the traditional $q$-state mean-field Potts model
emphasizing some points not often stressed in literature.
The model is defined through the following Hamiltonian built on the
fully connected (or complete) graph
\begin{eqnarray}
\label{HgPotts}
H=-\frac{J}{N}\sum_{i<j}\delta(\sigma_i,\sigma_j),
\end{eqnarray}
where $\delta(\sigma,\sigma')$ is the Kronecker delta function.
Let us rewrite $H$ as (up to terms negligible for $N\to\infty$)
\begin{eqnarray}
\label{HgPotts00}
H=-\frac{J}{N}\sum_\sigma \left[\sum_{i}\delta(\sigma_i,\sigma)\right]^2.
\end{eqnarray} 
From Eq. (\ref{HgPotts00}) we see that, if $J>0$, by introducing $q$ independent Gaussian variables, $x_\sigma$, $\sigma=1,\ldots,q$, we
evaluate the partition function, $Z= \sum_{\{\sigma_i\}}\exp(-\beta H(\{\sigma_i\}))$, as
\begin{eqnarray}
\label{Z1Potts0}
Z\propto \int \prod_{\sigma=1}^{q} d x_\sigma ~ e^{-N\left[\sum_\sigma \frac{\beta J x_\sigma^2}{2}
-\log\left(\sum_\sigma e^{\beta J x_{\sigma}}\right)
\right]}.
\end{eqnarray}  
From Eq. (\ref{Z1Potts0}), for $N\to\infty$, by saddle point, 
we get immediately
the following system of equations
\begin{eqnarray}
\label{Potts0t}
x_{\sigma}=\frac{e^{\beta J x_\sigma}}{\sum_{\sigma'}e^{\beta J x_{\sigma'}}}, \quad \sigma=1,\ldots,q,
\end{eqnarray}  
while the free energy density $f$ is given by
\begin{eqnarray}
\label{Pottsf0t}
\beta f=-\log\left(\sum_\sigma e^{\beta J x_{\sigma}}\right)+\sum_\sigma \frac{\beta J x_\sigma^2}{2},
\end{eqnarray}  
to be evaluated in correspondence of the solution of the 
system (\ref{Potts0t}).
We remind that 
the term represented by (\ref{Pottsf0t}) alone, where the $x_\sigma$'s are meant as free variables, is called
Landau free energy. For each $\sigma$, $x_\sigma$ coincides
with the thermal average $\sum_\sigma\exp(-\beta H)\sum_i\delta_{\sigma_i,\sigma}/N$,
\textit{i.e.}, the probability to find any spin in the state $\sigma$.
Eqs. (\ref{Potts0t}) are symmetric under permutation of the components
$(x_1,\ldots,x_q)$.
All the possible solutions of Eqs. (\ref{Potts0t}) can be found 
by setting $q-1$ components equal to each other and solving one
single equation.
If $(i_1,\ldots,i_q)$ is any permutation of $(1,\ldots,q)$, then we set
\begin{eqnarray}
\label{UA}
x_{i_1}=x, \quad x_{i_j}=y, \quad j=2,\ldots,q
\end{eqnarray}  
where $y=(1-x)/(q-1)$, and $x$ satisfies the equation 
\begin{eqnarray}
\label{UA1}
x=\frac{1}{1+(q-1)\exp\left[\frac{\beta J(1-qx)}{q-1}\right]}.
\end{eqnarray}  
Eqs. (\ref{Pottsf0t})-(\ref{UA1}) give rise to a well known phase transition scenario \cite{Wu}:
a second-order mean-field Ising phase transition sets up only for $q=2$, while for any 
$q\geq 3$ there is a first-order phase transition at the critical value (Fig.~\ref{fig1}):
\begin{eqnarray}
\label{tc}
\beta^{(\mathrm{F.O.})}_c J = \frac{2(q-1)}{q-2}\log(q-1).
\end{eqnarray}  
It is important to remind, however, that
besides the solution giving rise to the first-order phase transition,
and the trivial symmetric solution $x_1=\ldots=x_q=1/q$, which, in general,
can be leading, metastable or unstable, for $q>2$
Eqs. (\ref{Pottsf0t})-(\ref{UA1}) have
also a non trivial unstable solution giving rise to a second order phase
transition which sets in at the following critical point 
\begin{eqnarray}
\label{tc1}
\beta^{(\mathrm{S.O.~unstable})}_c J =q.
\end{eqnarray}  
Eq. (\ref{tc1}) establishes the temperature below which
the symmetric solution is no longer metastable or stable.
As we will see later, Eq. (\ref{tc1}) has a stable counterpart when $J<0$ and 
a discrete-time dynamics is considered.

\begin{figure}[thb]
\includegraphics[scale=0.35]{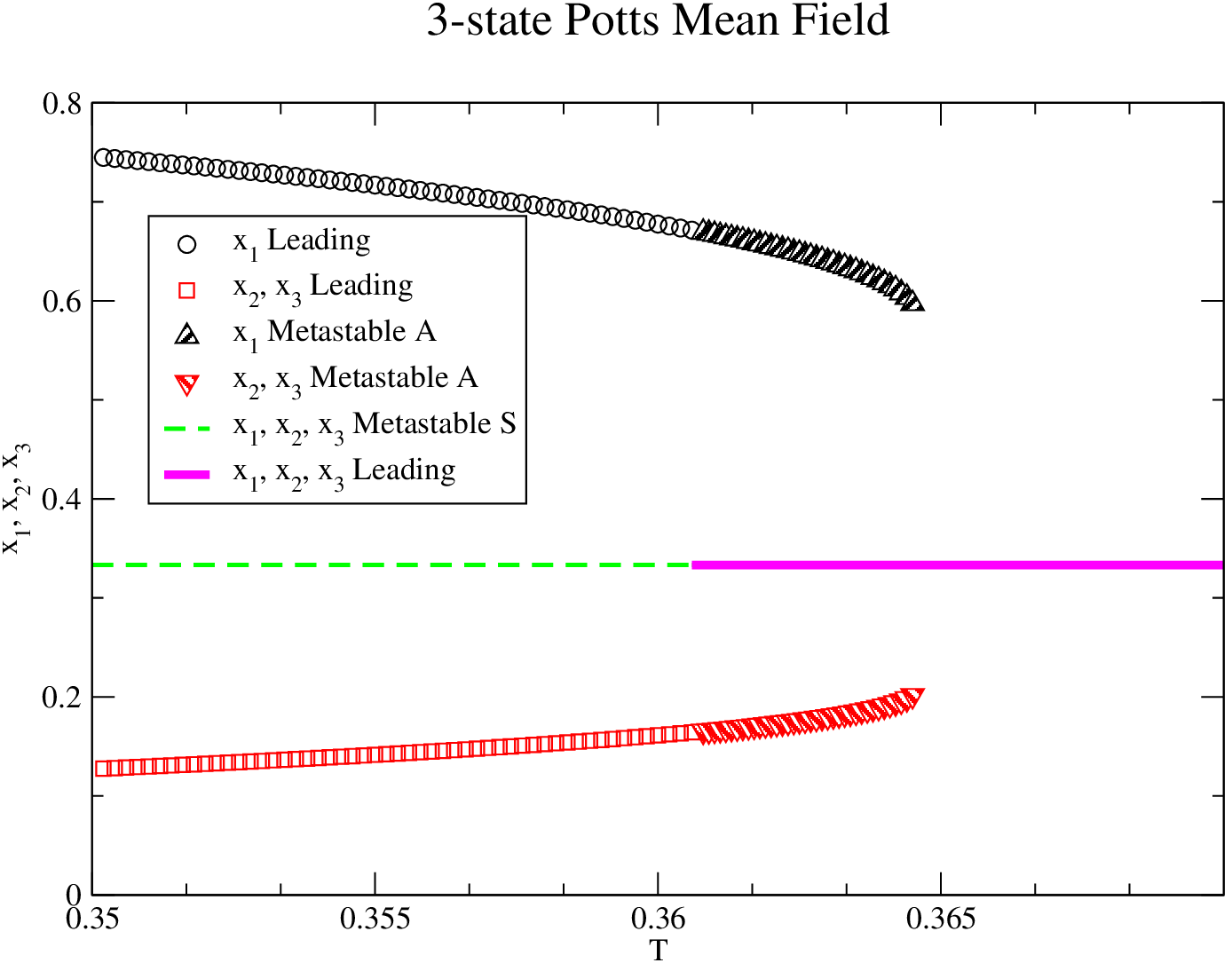}
\caption{(Color online) Enlargement of Fig. (\ref{fig1}). Notice, on this scale, 
the region of the metastable A (non symmetric) solution in a range of temperatures
larger than $T^{(\mathrm{F.O.})}_c=0.3607$.}
\label{fig2}
\end{figure}

In Figs. (\ref{fig1}) and (\ref{fig2}) we plot all the possible solutions of Eqs. (\ref{UA1}) and 
study their nature according to the free energy density (\ref{Pottsf0t}).
{
Note that a solution of Eqs. (\ref{Potts0t}) is: leading if it is the absolute minimum for $f$; 
metastable if it is a local minimum for $f$;
or unstable when is not a minimum for $f$ 
(\textit{i.e.}, at least one eigenvalue of the Hessian of the Landau free energy (\ref{Pottsf0t}) is not positive).  
}

\vspace{-0.8cm}
\section{Equilibrium versus dynamics and the antiferromagnetic case}
Above we have analyzed only the ferromagnetic case $J>0$.
For the antiferromagnetic case $J<0$, 
it is possible to prove that the system remains in the trivial symmetric
state \cite{Contucci}. 
By summarizing, at the Gibbs-Boltzmann equilibrium, for $q\geq 3$
the system undergoes a first-order phase transition only in
the ferromagnetic case $J>0$, while in the antiferromagnetic case $J<0$
there is no phase transition.

A similar situation applies when one considers instead 
a dynamical approach, like the continuous-time Glauber dynamics \cite{Glauber}.
The Glauber dynamics is a single spin-flip dynamics
governed by the following master Eq.
\begin{eqnarray}
\label{Master}
&& \frac{d p(\sigma_1,\ldots,\sigma_N;t)}{\alpha dt}=
\\ && 
-\sum_i\sum_{\sigma^{'}_i}w(\sigma_i\to\sigma^{'}_i)p(\sigma_1,\ldots,\sigma_N;t)
+ \nonumber \\ && 
\sum_i\sum_{\sigma^{'}_i}w(\sigma^{'}_i\to\sigma_i)
p(\sigma_1,\ldots,\sigma_{i-1},\sigma^{'}_i,\sigma_{i+1},\ldots\sigma_N;t),
\nonumber
\end{eqnarray}  
where $p(\sigma_1,\ldots,\sigma_N;t)$ is the probability that the system
is in the configuration $\sigma_1,\ldots,\sigma_N$ at time $t$,
$w(\sigma\to\sigma^{'})$ are transition rate probabilities due to interaction with other spins,
and $\alpha/q$ is the rate per unit time at which,
due to the interaction with an heat reservoir,
a free Potts spin makes transitions from either state to
the other $q-1$. 
By imposing that, for $t\to\infty$, the system reaches equilibrium
according to the Gibbs-Boltzmann probability 
$p(\sigma_1,\ldots,\sigma_N;t)\propto \exp(-\beta H)$, with $H$ given
by Eq. (\ref{HgPotts}), and by using the strong law of large numbers,
it is easy to see that the transition rates must satisfy the 
following functional Eq. (also known as detailed-balance)
\begin{eqnarray}
\label{Rates}
\sum_{\sigma}e^{\beta J x_\sigma}w(\sigma\to\sigma^{'})=e^{\beta J x_{\sigma^{'}}},
\end{eqnarray}  
where, as in the previous Section, $x_\sigma$ is
the probability to find any spin in the state $\sigma$ when $t\to\infty$.
We will keep using the same symbol to indicate the probability
to find any spin $\sigma_i$ in the state $\sigma$ also for finite $t$:
$p(\sigma_i=\sigma;t)=x_\sigma(t)$.
Taking into account the normalization condition 
$\sum_{\sigma^{'}} w(\sigma\to\sigma^{'})=1$,
we will consider the following natural solution of Eq. (\ref{Rates}) \cite{Mendes}
\begin{eqnarray}
\label{Rates1}
w(\sigma\to\sigma^{'})=\frac{e^{\beta J x_{\sigma^{'}}}}
{\sum_{\sigma^{''}}e^{\beta J x_{\sigma^{''}}}}.
\end{eqnarray}  
Rates (\ref{Rates1}) have the remarkable feature, typical of mean-field models,
that they depend only on the final states. This fact allows
us to immediately derive the dynamical Eqs. governing the
evolution of the $x_{\sigma}(t)$'s (also known as reduced dynamics \cite{Glauber}). 
By plugging Eqs. (\ref{Rates1}) in (\ref{Master}) we get
the following $q$-coupled (of which $q-1$ are independent) differential Eqs. of the order parameters $x_{\sigma}(t)$
\begin{eqnarray}
\label{PGlauber0}
\frac{d x_{\sigma}(t)}{\alpha dt}=-x_{\sigma}(t)+
\frac{e^{\beta J x_\sigma(t)}}{\sum_{\sigma'}e^{\beta J x_{\sigma'}(t)}},
\end{eqnarray}  
Eqs. (\ref{PGlauber0}) can be more simply derived also from the time-dependent
Ginzburg-Landau Eqs. $dx_\sigma/dt \propto \partial f(\left\{x_\sigma'\right\})/\partial x_\sigma$,
as it has been done to study the ferromagnetic Potts model in the presence of 
an oscillating external field \cite{Kinoshita}. We here, however, wanted
to start from a microscopic probabilistic approach.
Of course, in the stationary limit $t\to\infty$,  
Eqs. (\ref{PGlauber0}) are equivalent
to the self-consistent Eqs. (\ref{Potts0t}). Furthermore,
among the stationary solutions of Eqs. (\ref{PGlauber0}),
the dynamically stable ones coincide with the minima of the 
Landau density free-energy at equilibrium (\ref{Pottsf0t}).
And this holds for both the cases $J>0$ and $J<0$.
As a consequence, apart from an initial transient, 
the Potts model at Gibbs-Boltzmann equilibrium, or along the continuous Glauber dynamic,
do not present a significant difference.

Note that, the lhs of Eqs. (\ref{PGlauber0}) and the rate $\alpha$, that determine
how fast the system reaches equilibrium, 
are due to the interaction of the
spins with the medium environment, \textit{e.g.} heat, 
while the rhs of Eqs. (\ref{PGlauber0}) stems from the interactions of
neighboring spins at a given $\beta$.

In the place of the above continuous-time dynamics, we introduce now 
a discrete-time dynamics taking place at the 
discrete times $t=0,1,2,\ldots,$ by the following Eqs. 
\begin{eqnarray}
\label{PGlauberD0}
&& \frac{x_{\sigma}(t+1)-x_{\sigma}(t)}{\alpha}= 
-x_{\sigma}(t)+
\frac{e^{\beta J x_\sigma}}{\sum_{\sigma'}e^{\beta J x_{\sigma'}}}.
\end{eqnarray}   
Eqs. (\ref{PGlauberD0}) can be derived from a discrete-time master Eq. 
in close analogy to the continuous case, the difference being
the assumption that we consider only discrete times and then finite
difference probabilities. 
{
Furthermore, we need to work with global transition rate probabilities defined
in the full $q^N$-dimensional space of the $N$ spins.
More precisely, if we introduce the 
spin vector $\bm{\sigma}=(\sigma_1,\ldots,\sigma_N)$, and the 
associated probability vector $p(\bm{\sigma})$, 
instead of Eq. (\ref{Master}), we have
\begin{eqnarray}
\label{Master1}
&& \frac{p(\bm{\sigma};t+1)-p(\bm{\sigma};t)}{\alpha}=
-\sum_{\bm{\sigma}^{'}}p(\bm{\sigma};t)W(\bm{\sigma}\to\bm{\sigma}^{'}) ~~~~ \\ && 
+\sum_{\bm{\sigma}^{'}}p(\bm{\sigma}^{'};t)W(\bm{\sigma}^{'}\to\bm{\sigma}),
\nonumber
\end{eqnarray}  
where we have introduced the global transition rates in terms of the local weights (\ref{Rates1}):
$W(\bm{\sigma}\to\bm{\sigma}^{'})=\prod_{i=1}^N w(\sigma_i\to\sigma_i^{'})$
As will be made clear later, for the model to be consistent, unlike the continuous case,
we must impose the boundary $\alpha\leq 1$.
The difference between Eqs. (\ref{PGlauber0}) and Eqs. (\ref{PGlauberD0})
is that in Eqs. (\ref{PGlauber0}) the medium acts continuously in time, 
while in Eqs. (\ref{PGlauberD0}) the medium acts only at discrete times and,
between one time and the next one, the spins do not move 
(alternatively, we can can see the spin as frozen).
Whereas only Eqs. (\ref{PGlauber0}) can represent some description
of a system of physical particles each other interacting via a physical medium, 
Eqs. (\ref{PGlauberD0}) can represent a system of agents 
which interact via, \textit{e.g.}, exchange of information
taking place at discrete random times~\footnote{
For simplicity, in Eqs. (\ref{PGlauberD0}) we have assumed $t=0,1,2,\ldots$.
Note, however, that the modification to include arbitrary random times is elementary since 
the rhs of Eqs. (\ref{PGlauberD0})
involves only an implicit dependence on the time via the $x_i(t)$'s. As a consequence,
the probability distribution for the random times does not affect 
the behavior of the system. It affects only the times at which the changes of the $x_i(t)$'s
may be observed.}, as in fact occurs in the actual world,
especially in social or economical contexts.

When $J>0$, the difference between Eqs. (\ref{PGlauber0}) and Eqs. (\ref{PGlauberD0})
is not dramatic as, for $t\to\infty$, their solutions both tend to the the same $\alpha$-independent
stable point-like stationary solutions of Eqs. (\ref{Potts0t}) (see Fig. (\ref{fig6})). 
However, when $J<0$, for $t\to\infty$, whereas the solution of Eqs. (\ref{PGlauber0})
tends simply to the trivial symmetric one,
the solution of Eqs. (\ref{PGlauberD0}) tends, for $t\to\infty$, 
to a period-2 stable trajectory, \textit{i.e.}, the solution oscillates
between two values (see Figs. (\ref{fig7}) and (\ref{fig8})). Furthermore, with respect to the
temperature, this solution undergoes a second-order phase transition
belonging to the mean-field Ising universality class (as shown in Fig.~(\ref{fig5}),
which represents the case $\alpha=1$). By changing $\alpha$, the critical behavior
remains unchanged (and the curves are similar to those shown in Fig.~(\ref{fig5})), 
while the critical temperature turns out to be a growing function of $\alpha$.
Analytically, the mechanism can be understood and the critical temperature evaluated as follows.
Let us look again for solutions in the form (\ref{UA}).
From Eqs. (\ref{PGlauberD0}) we have
\begin{eqnarray}
\label{xt1}
x(t+1)=x(t)(1-\alpha)+\frac{\alpha}{1+(q-1)
\exp\left(\beta J\frac{1-qx(t)}{q-1}\right)}.
\end{eqnarray}  
Of course, the stationary solution of Eq. (\ref{xt1}) that we get by imposing $x(t+1)=x(t)=x$,
coincides with Eq. (\ref{UA1}), so that, in particular,
we already know that such stationary solutions
do not undergo any phase transition when $J~<~0$.
It is worth to remind that, in general, even tough a stationary Eq. has a solution, this does not
necessary mean that the system reaches equilibrium. Eq. (\ref{xt1}) for $J<0$
represents such a case: there exists the trivial stationary solution $x_\sigma=1/q$
but, nevertheless, the system never reaches equilibrium. 
Let us consider now the next 
\vspace{-0.4cm} 
time evolution. From Eq. (\ref{xt1}) we have
\begin{eqnarray}
\label{xt2}
&& x(t+2)=x(t)(1-\alpha)^2+\frac{\alpha(1-\alpha)}{1+(q-1)\exp\left(\beta J\frac{1-qx(t)}{q-1}\right)} \nonumber \\
&& + \alpha \left[1+(q-1)\exp\left(\beta J \right. \right. \nonumber \\ 
&& \times \left. \left. \frac{1-q\left[x(t)(1-\alpha)+
\frac{\alpha}{1+(q-1)\exp\left(\beta J\frac{1-qx(t)}{q-1}\right)}\right]}{q-1}\right)\right]^{-1}
\end{eqnarray}   

Let us analyze the stationary solutions of Eq. (\ref{xt2}) that we get
by imposing $x(t+2)=x(t)=x$, \textit{i.e.}, the period-2 stable trajectories. 
By writing $x=1/q-\epsilon$, 
and expanding Eq. (\ref{xt2}) at the first order in $\epsilon$, it is not
difficult to see that the solution $\epsilon=0$ becomes unstable when,
for given $q$ and $\alpha$,
the temperature satisfies the following Eq.
\begin{eqnarray}
\label{Tca0}
\alpha(\beta J)^2+2q(1-\alpha)\beta J -q^2(2-\alpha)=0.
\end{eqnarray}   
As expected, when $J>0$, the only real root of Eq. (\ref{Tca0}) coincides with
the critical (unstable) temperature of the equilibrium case Eq. (\ref{tc1}).
However, when $J<0$, the only real root of Eq. (\ref{Tca0}) is 
\begin{eqnarray}
\label{Tca}
\beta^{(\mathrm{S.O.})}_c J=q\left(1-\frac{2}{\alpha}\right), \quad \alpha\leq 1.
\end{eqnarray}   
Eq. (\ref{Tca}) says that the critical temperature is a growing function of $\alpha$. Notice
that $\alpha=1$ returns the same value given by Eq. (\ref{tc1}) with $J$ replaced by $|J|$
but, now, at low $T$, the states live on a subspace that oscillates between both the space of the states
featured by the stable first-order transition that we had for $J>0$, and the space of the states of the unstable second-order
transition that we had for $J>0$ (compare Figs. (\ref{fig1}) and (\ref{fig5})).

Finally, by expanding Eq. (\ref{xt2}) further, up to the third order in $\epsilon$,
and by using Eq. (\ref{Tca}), we obtain
\begin{eqnarray}
\label{Crit}
A\tau \epsilon+(q-2)B\tau\epsilon^2+C\epsilon^3+\mathop{O}(\epsilon^4)=0,
\end{eqnarray}   
where $\tau=(T^{(\mathrm{S.O.})}_c-T)/T^{(\mathrm{S.O.})}_c$, and $A$, $B$, and $C$ are constants that do not depend on $\tau$.
From (\ref{Crit}) it follows that the critical behavior is Ising-like for any $\alpha$:
$\epsilon\propto \tau^{1/2}$ (see enlargement of Fig. (\ref{fig5})).
Notice that in the second term of the lhs of Eq. (\ref{Crit}) we made explicit a dependence on the factor $q-2$ so that,
the Ising case $q=2$, where the second term cancels exactly, is included as a special case.

We come now back to the necessary condition $\alpha\leq 1$.
In fact, unlike the continuous case, in the lhs of the master Eq. (\ref{Master1}) we have a difference
of probabilities. Whereas the derivatives $dp/dt$ can take any real
value, the finite difference of probabilities $p(t+1)-p(t)$ that
appear in Eqs. (\ref{Master1}), must lie in the interval [-1,1]. 
Or, in other words, we have to ensure
that, given that $p(t)$ is a probability, $p(t+1)$ is also a probability.
Note that, for any real $\alpha$, both Eqs. (\ref{Master}) or (\ref{Master1}) give rise
to normalized $p(\sigma_1,\ldots,\sigma_N;t)$'s at any $t$, but, in general, it may happen
that the $p(t)$'s can take values outside the probability range $[0,1]$.
{
Let us rewrite Eq. (\ref{Master1}) in compact matrix notation:
\begin{eqnarray}
\label{Master1M}
\bm{p}(t+1)=\bm{p}(t)\left[\alpha \bm{W}+(1-\alpha)\bm{1}\right].
\end{eqnarray}  
From Eq. (\ref{Master1M}) we see that the matrix appearing to the right hand side is a stochastic
matrix (\textit{i.e.} {all the entries are positive and the rows are normalized to 1}) when $\alpha<1$ and,
as a consequence the Markov chain $\bm{p}(t)$ is a well defined probability.
}
\begin{figure}[thb]
\includegraphics[scale=0.35]{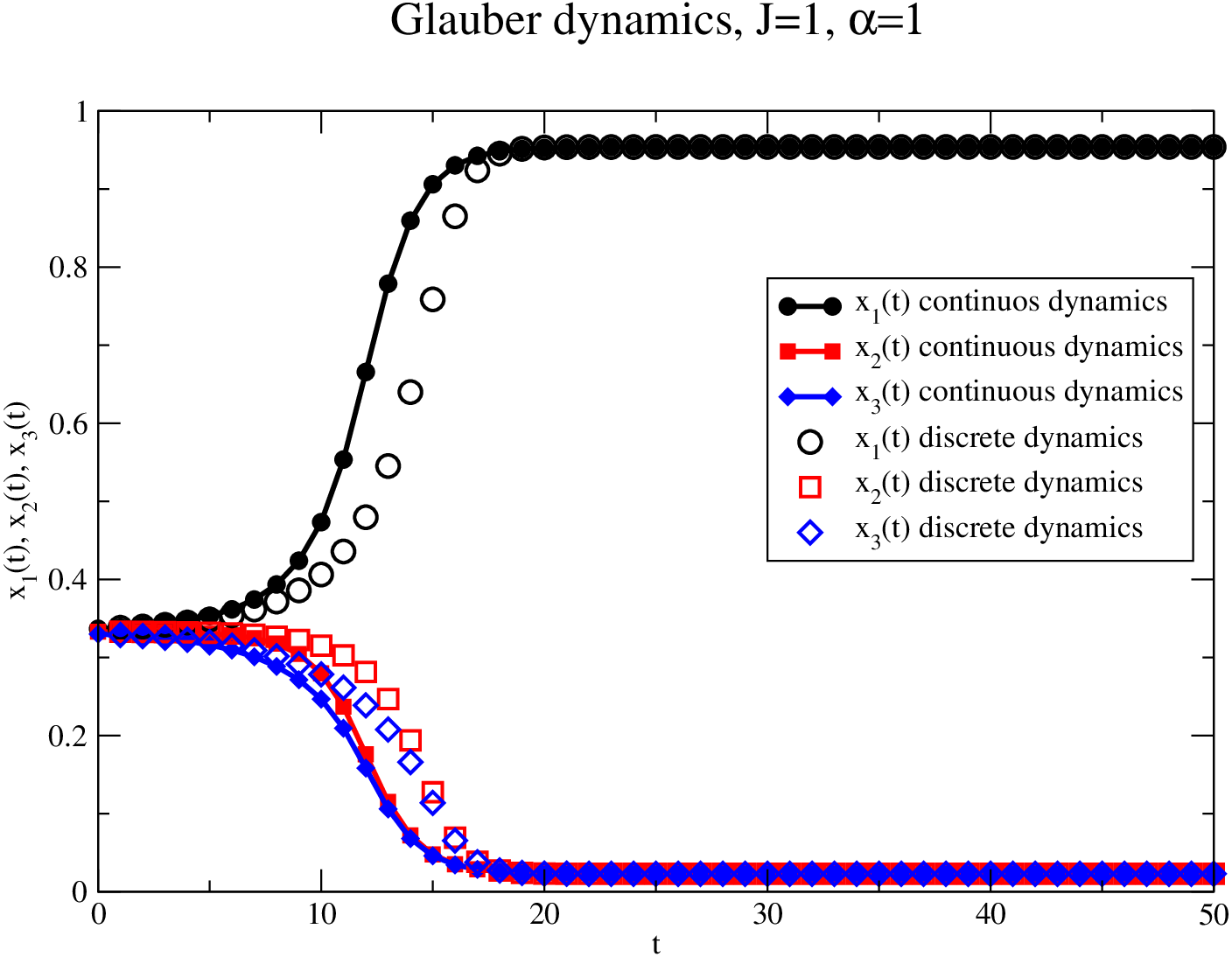}
\caption{(Color online) Continuous- and discrete-time Glauber dynamics
for a 3-state Potts model with ferromagnetic coupling $J=1$ and $\alpha=1$. 
The continuous and discrete case are solutions of Eqs. (\ref{PGlauber0}) and
(\ref{PGlauberD0}), respectively, with initial conditions
close to the symmetric solution: 
{
$x_1(0)=0.3369$, $x_2(0)=0.3331$, $x_3(0)=0.33$.  
}
The chosen temperature is $T/J=1/4$, which
is lower than the first-order critical temperature of this case, $T^{(\mathrm{F.O.})}_c/J=0.3607$.} 
\label{fig6}
\end{figure}

\begin{figure}[thb]
\includegraphics[scale=0.35]{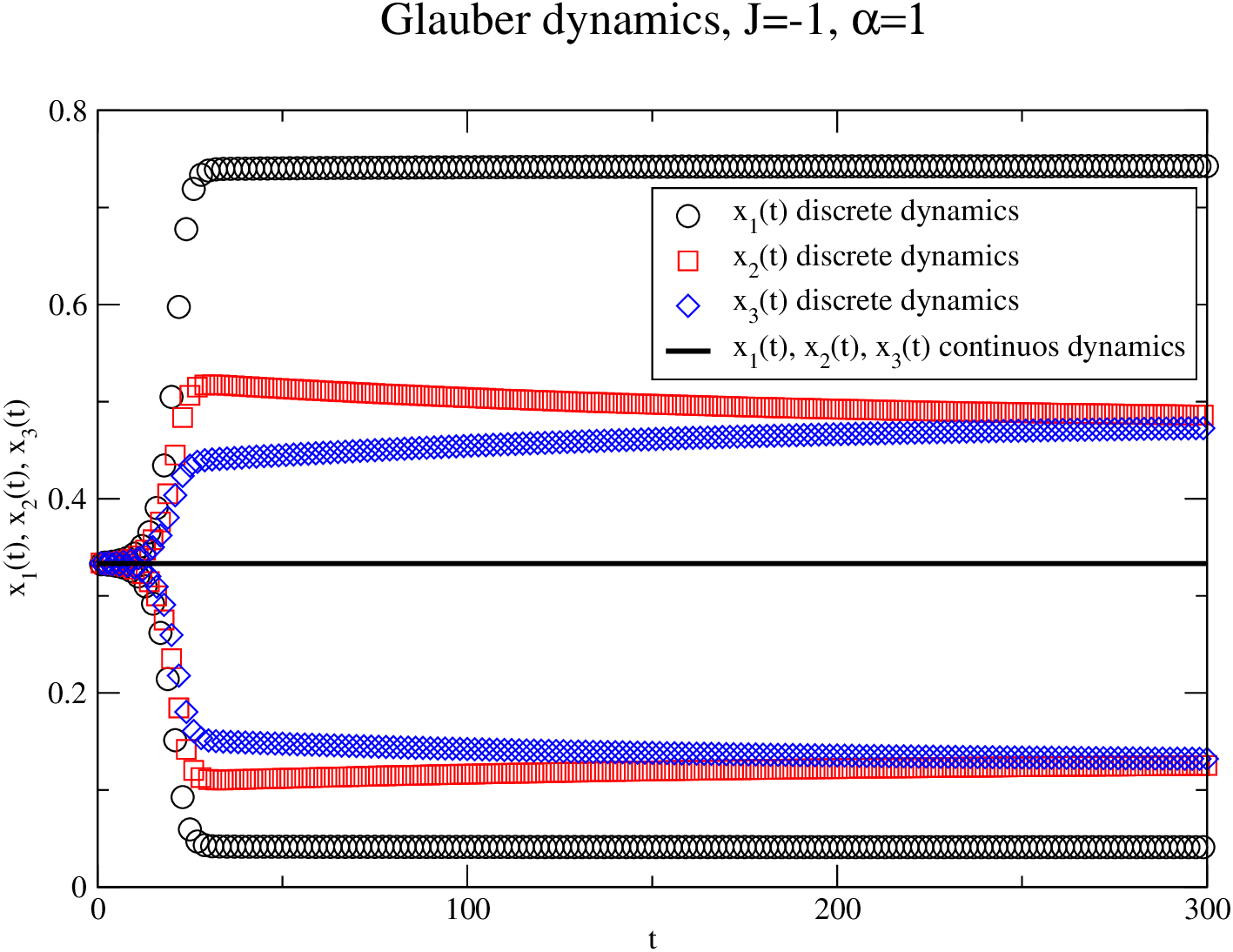}
\caption{(Color online) Continuous- and discrete-time Glauber dynamics
for a 3-state Potts model with anti-ferromagnetic coupling $J=-1$ and $\alpha=1$. 
The continuous and discrete case are the solutions of Eqs. (\ref{PGlauber0}) and
(\ref{PGlauberD0}), respectively, with initial conditions as in Fig. \ref{fig6}.
The chosen temperature is $T/J=1/4$, which
is lower than the critical temperature of this case, $T^{(\mathrm{S.O.})}_c/|J|=1/3$, where a second-order
phase transition takes place. 
{
Note that the system never reaches point-like equilibrium
(see Fig. (\ref{fig8})).
} 
}
\label{fig7}
\end{figure}

\begin{figure}[thb]
\includegraphics[scale=0.35]{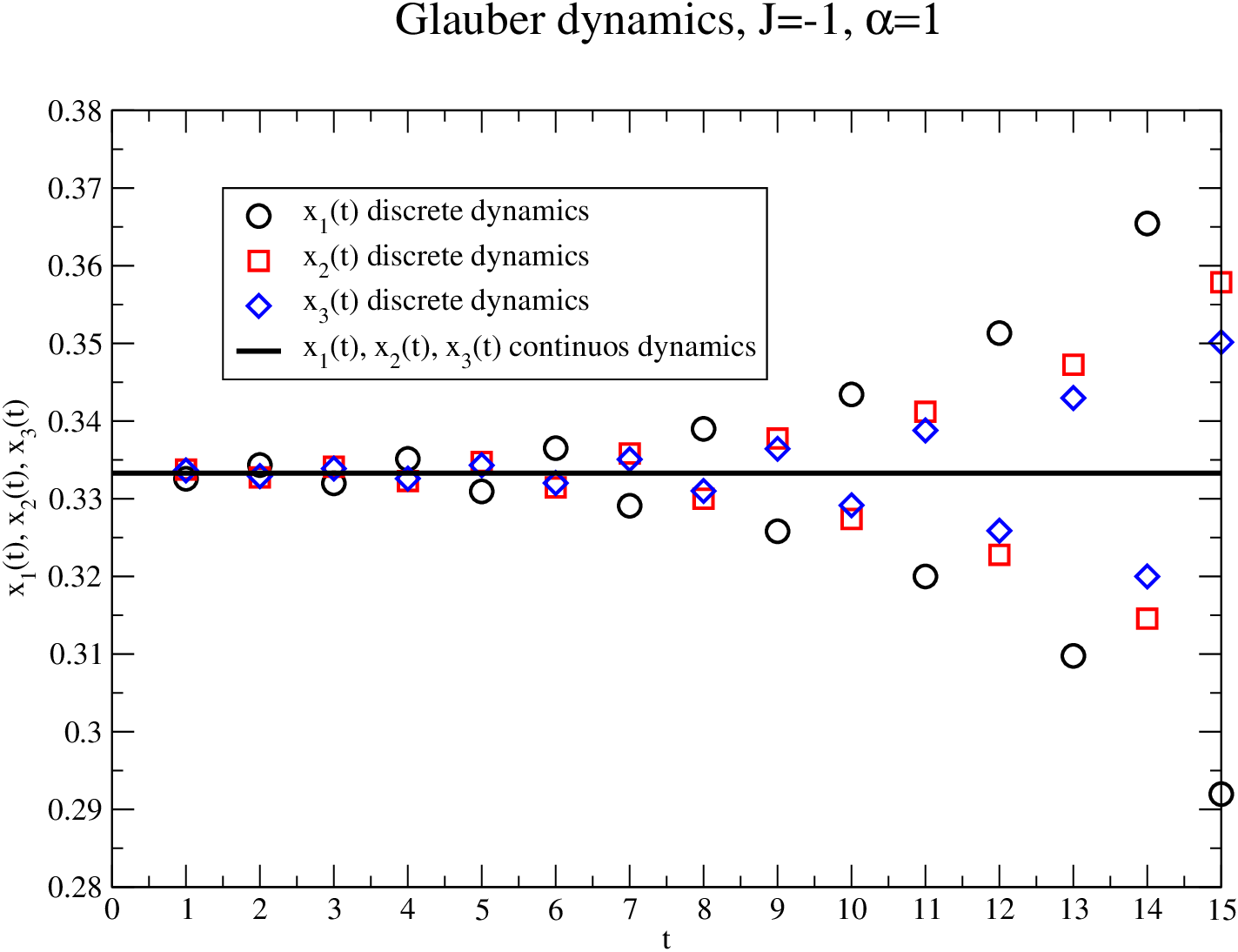}
\caption{(Color online) Enlargement of Fig. \ref{fig7}. 
Notice, on this scale, the oscillating behavior of each component.}
\label{fig8}
\end{figure}
  
\begin{figure}[thb]
\includegraphics[scale=0.35]{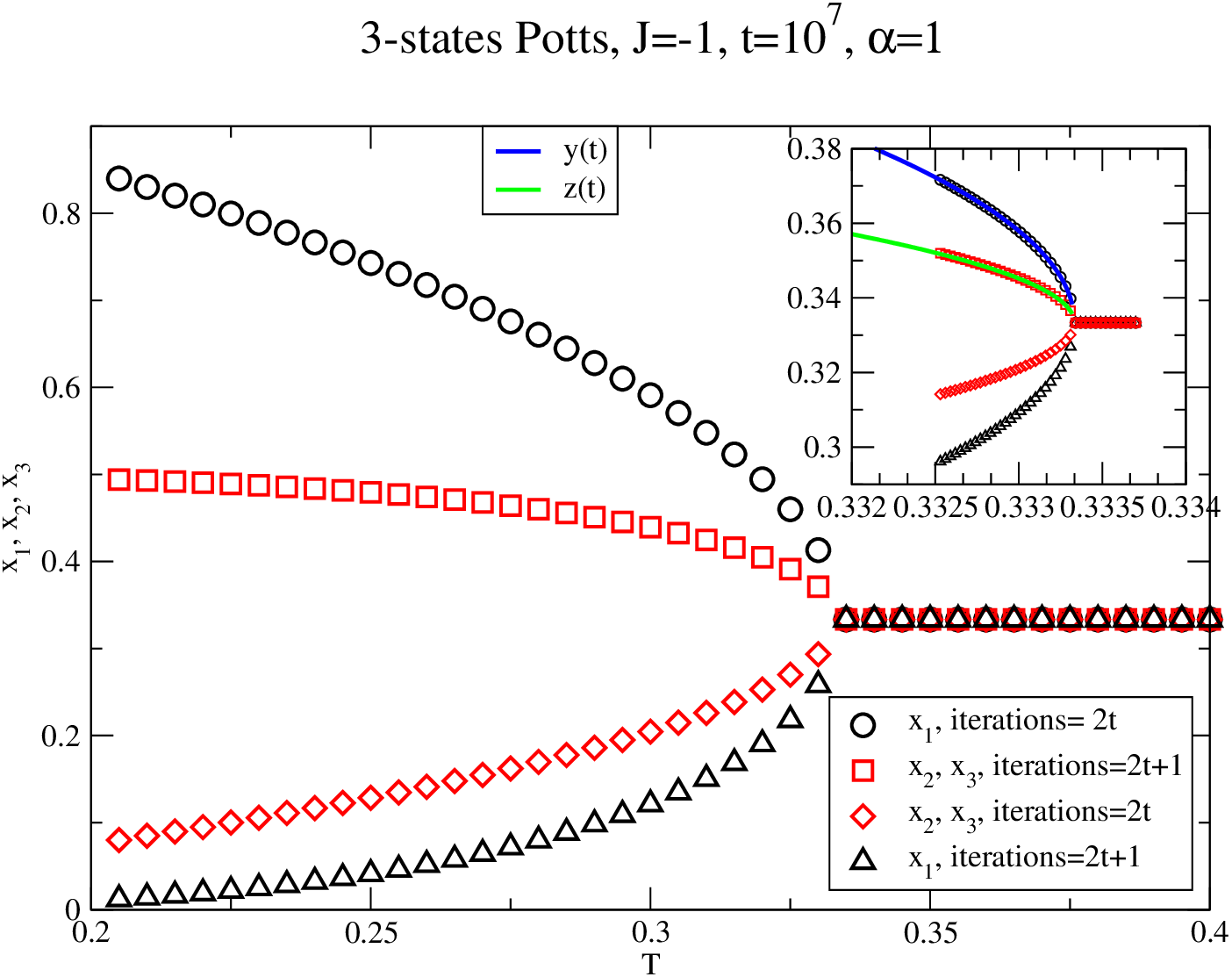}
\caption{(Color online) Magnetizations of the period-2 attractor 
of Eqs. (\ref{PGlauberD0}) for the case $q=3$, $J=-1$ and $\alpha=1$. 
{
Once reached the attractor (for large enough discrete times $t$), 
}
each component $x_i$ oscillates periodically 
taking the two values $x_i(2t)$ and $x_i(2t+1)$
(in this figure only the case $x_1>x_2=x_3$ for $t$ even, and 
$x_1<x_2=x_3$ for $t$ odd, are shown).
Inset: enlargement around the transition point where, for $T\leq T^{(\mathrm{S.O.})}_c$, 
we plot also the matching functions $y(T)=1/3+1.35 (T^{(\mathrm{S.O.})}_c-T)^{1/2}$
and $z(T)=1/3+0.65 (T^{(\mathrm{S.O.})}_c-T)^{1/2}$, with $T^{(\mathrm{S.O.})}_c=1/3$, 
that confirm the Ising-like critical behavior.} 
\label{fig5}
\end{figure}

\section{Conclusions}
We have shown that anti-ferromagnetic models
that evolve along a discrete-time dynamics, as opposed to
a continuous-time dynamics, reach period-2 stable trajectories
where, after an initial transient, the order parameter oscillates between two values
each of which undergoes a phase transition. Quite unexpectedly,
the nature of such a phase transition is second-order, even though
the only possible phase transitions of the Gibbs-Boltzmann equilibrium
case were first-order. We have analyzed here the mean-field $q$-state Potts case
in detail, 
but we believe this scenario being quite typical and robust with
respect to the model definition.
The results of our analysis shed some light to real-world models.
Many real-world models, like social networks, population growths, 
economic networks, etc..., are in fact characterized by
random interactions that take place not continuously, but only at certain discrete
random times. Furthermore, a portion of these interactions are
often unfriendly, \textit{i.e.}, anti-ferromagnetic. Therefore,
at least at the mean-field level, and in the absence of quenched disorder,
that would involve a glassy behavior \cite{Parisi}, such real-world models are expected
to reach a stable oscillating behavior and undergo a second-order phase transition
for each component of the order parameter.

\begin{acknowledgments}
Work supported 
by Grant IIUM EDW B 11-159-0637 and
``Finanziamento per la Ricerca Scientifica - Sapienza Universit\`a di Roma - prot.
C26A1179MF''. We thank F. Cesi and F. Ricci-Tersenghi for useful discussions.
\end{acknowledgments}



\begin{thebibliography}{10}

\bibitem{Wu} F. Y. Wu, Rev. Mod. Phys., \textbf{54}, 235 (1982).


\bibitem{Fortuin}
C. M. Fortuin and P. W. Kasteleyn, 1972, Physica \textbf{57}, 536 (1972).

\bibitem{Zecchina}
O. C. Martin, R\'emi Monasson, R. Zecchina,
Theor. Comp. Science \textbf{265} 3 (2001).

\bibitem{MezardC}
M. M\'e zard, T. Mora, and R. Zecchina, Phys. Rev. Lett.
\textbf{94}, 197205 (2005).

\bibitem{Reichardt}
J. Reichardt, and S. Bornholdt, Phys. Rev. Lett. \textbf{93},
218701 (2004).

\bibitem{Review} S.N. Dorogovtsev, A.V. Goltsev, J.F.F. Mendes, Rev. Mod. Phys. \textbf{80}, 1275 (2008).

\bibitem{Contucci} However, a phase transition can take place
also for $J<0$ in finite connectivity graphs: 
P. Contucci, S. Dommers, C. Giardin\'a, S. Starr,
arXiv:1106.4714 (2012).

\bibitem{Glauber} R. J. Glauber, J. Math. Phys. \textbf{4}, 294 (1963).

{
\bibitem{Costeniuc} M. Costeniuc, R. S. Ellis, and Hugo Touchette,
J. Math. Phys. \textbf{46}, 063301 (2005).
}

\bibitem{Mendes} We note the existence of other
solutions leading to different Eqs. for the evolution of the
order parameter: 
J. F. F. Mendes and E. J. S. Lage,
J. Stat. Phys., \textbf{64}, 653 (1991).

\bibitem{Kinoshita}
T. Kinoshita \textit{et al.},
Int. Conf. on Magnetism IOP Publ.,
Journal of Physics: Conference Series \textbf{200} 022026 (2010).

\bibitem{Parisi} M. Mezard, G. Parisi, M. A. Virasoro 
\textit{Spin Glass Theory and Beyond} (World Scientific) (1987).

\bibitem{Synchro}
W. A Little, Math. Biosci. \textbf{19}, 101–20 (1974);
P Peretto, Biol. Cybern. \textbf{50}, 51 (1984);
J. L. Lebowitz \textit{et al.}, J. Stat. Phys. \textbf{59}, 117 (1990);
N. S. Skantzos and A C. C. Coolen, J. Phys. A: Math. Gen. \textbf{33}, 1841 (2000).

\end{thebibliography}
\end{document}